\pdfoutput=1
\documentclass[twoside]{IEEEtran}
\usepackage[caption=false,font=footnotesize]{subfig}
\usepackage{xcolor}    
\usepackage{amsmath}   
\allowdisplaybreaks[4]
\usepackage{amssymb}   
\usepackage{amsfonts}  
\usepackage{mathrsfs}  
\usepackage{array,bm}  
\usepackage{textcomp}  
\usepackage{stfloats}  
\usepackage{verbatim}  
\usepackage{graphicx}  
\usepackage{cite,url}  
\usepackage{booktabs}  
\usepackage{multirow}  
\usepackage{lettrine}  
\usepackage{hyperref}  
\usepackage{mathtools} 
\usepackage{orcidlink} 
\usepackage[ruled,linesnumbered]{algorithm2e} %
\usepackage{lipsum}
\graphicspath{{pic/}} 

\begin{document}

\newcommand{\tabincell}[2]{
  \begin{tabular}{@{}#1@{}}#2
\end{tabular}} %
\newtheorem{Property}{\it Property}
\newtheorem{Proposition}{\bf Proposition}
\newtheorem{remark}{Remark}
\newenvironment{Proof}{{\indent \it Proof:}}{\hfill $\blacksquare$\par}

\title{Flexible Cylindrical Array-Aided Secure Wireless Communications}

\author{
  Xiangyu Dong$^{\orcidlink{0009-0004-4606-2271}}$, 
  Ran Yang$^{\orcidlink{0000-0002-0435-7926}}$,
  Songjie Yang$^{\orcidlink{0000-0003-3130-4747}}$,
  Weidong Mei$^{\orcidlink{0000-0002-8113-4283}}$,~\IEEEmembership{Member,~IEEE},\\
  Lipeng Zhu$^{\orcidlink{0000-0002-7587-8876}}$,~\IEEEmembership{Senior Member,~IEEE},
  Yue Xiu$^{\orcidlink{0000-0002-8433-6497}}$,~\IEEEmembership{Member,~IEEE},
  Zhongpei Zhang$^{\orcidlink{0000-0003-2772-9937}}$,~\IEEEmembership{Member,~IEEE}

  \thanks{
  Xiangyu Dong, Weidong Mei, Ran Yang, Songjie Yang, Yue Xiu and Zhongpei Zhang are with the National Key Laboratory of Wireless Communications, University of Electronic Science and Technology of China, Chengdu 611731, China. (e-mail: meiw@uestc.edu.cn, zhangzp@uestc.edu.cn).
  
  Lipeng Zhu is with the State Key Laboratory of CNS/ATM and the School of Interdisciplinary Science, Beijing Institute of Technology, Beijing 100081, China (E-mail: zhulp@bit.edu.cn).}
}
\maketitle
\begin{abstract}
  Flexible-geometry arrays based on movable antennas have shown considerable potential for improving wireless communication performance.
  In this letter, we investigate a multiuser multiple-input single-output (MU-MISO) downlink secure communication system aided by a flexible cylindrical array (FCLA) and artificial noise (AN), where each antenna element rotates along circular tracks while the circular slices move along a vertical axis.
  To guarantee transmission security, we aim to maximize the achievable sum rate at multiple legitimate information receivers by jointly optimizing transmit beamforming, AN covariance matrix, and antenna placement under secrecy constraints for an eavesdropper.
  While the resulting problem is intractable to solve, we develop a block coordinate descent (BCD)-based framework that combines the Lagrangian dual transform, tight semidefinite relaxation (SDR), and Nesterov-accelerated projected gradient descent (PGD).
  Numerical results show that the proposed algorithm converges rapidly and achieves significant sum-rate gains over benchmark schemes by exploiting the geometry flexibility of the array.
\end{abstract}


\section{Introduction}

\IEEEPARstart{I}n the forthcoming era of sixth-generation (6G) networks, massive communication and ubiquitous connectivity are envisioned to support increasingly dense and heterogeneous devices.
This trend is especially evident in industrial Internet-of-Things (IoT) and embodied-intelligence systems, with large numbers of sensors, robots, and terminals requiring simultaneous access under stringent rate, reliability, and security requirements\cite{MW1}.
Consequently, wide area coverage and secure transmission become tightly coupled design goals in multiuser systems\cite{Ma11}. 

Circular and cylindrical arrays present an advantageous topology, with rotational symmetry supporting uniform $360^\circ$ azimuth coverage and flexible spatial control across directions \cite{Tan2022ArrayTopologies}.
When users are widely distributed over azimuth, cylindrical arrays are envisioned to reduce the effective-aperture loss and improve the spatial separability of large-incidence-angle links, compared to fixed linear or planar arrays \cite{ZhouCong1,XIU4,CLA7,Wu2023UCA,Wu2024UCA,ISAC_FCLA}.   
For instance, the authors in \cite{Wu2023UCA} proposed to exploit uniform circular arrays (UCAs) to achieve a wide service region. In \cite{Wu2024UCA}, the authors studied beam-defocus effects in wideband millimeter-wave (mmWave) and terahertz (THz) UCA systems. In \cite{ISAC_FCLA}, the authors pioneered the deployment of UCAs in integrated sensing and communication (ISAC) systems to mitigate sidelobe interference.
Furthermore, the authors in \cite{CLA4} highlighted the performance advantages of rotationally symmetric arrays for wide-area user distributions in mmWave cellular networks.

While expanding transmission coverage, these geometric benefits have also drawn significant attention in enhancing communication security \cite{CLA1,XIU5,CLAS}. For instance, the authors in \cite{CLA1} exploited massive cylindrical arrays for physical layer security, improving secrecy capacity through directive transmission, jamming, and interference suppression. The public-security-oriented uncrewed aerial vehicle (UAV) detection problem was studied in \cite{CLAS} by leveraging a cylindrical array to enable omnidirectional scanning and flexible beam steering in cluttered environments. 
Nevertheless, these studies remain confined to fixed array geometries, which leads to limited secrecy performance gains.

Recently, movable antennas (MAs), also known as fluid antennas, introduce antenna-position adaptation as a new exploitable degree of freedom (DoF)\cite{MA1,zlp1,zlp2,MEI2,XIU1,XIU2}. Recent studies have already revealed their benefits in terms of secure communications\cite{zlp4}, including secrecy-rate gains with MA-inspired frequency-switching arrays\cite{ZhouCong2}, ergodic secrecy-rate improvement in multiple-input multiple-output multiple-antenna eavesdropper (MIMOME) systems\cite{XieLei}, secrecy gain with coexisting multicasting transmission\cite{MEI3}, and covert sum-rate enhancement in secure ISAC systems\cite{Ran1}. 
However, most existing studies still focus on linear or planar arrays, inevitably leading to reduced beamforming gains and user rates in wide-coverage scenarios.

In this letter, we investigate a flexible cylindrical array (FCLA)-assisted secure multiuser multiple-input single-output (MU-MISO) system, where a FCLA-equipped legitimate transmitter (Alice) communicates with multiple legitimate receivers (IRs) in the presence of an eavesdropper (Eve). In particular, we aim to maximize the sum rate at the IRs by jointly optimizing the transmit beamforming, artificial noise (AN) covariance matrix, and antenna placement subject to secrecy constraints for the Eve. The resulting optimization problem turns out to be intractable due to the highly non-linear mapping from the antenna positions to channel coefficients, as well as the tight coupling among optimization variables.
To deal with this issue, we develop a block coordinate descent (BCD) algorithm that combines the Lagrangian dual transform, tight semidefinite relaxation (SDR)-based beamforming and AN covariance optimization, and Nesterov-accelerated projected gradient descent (PGD) procedures. Simulation results demonstrate that the proposed scheme converges rapidly and significantly outperforms benchmark schemes in terms of system sum rate.
\section{System Model}\label{sys_model}
\vspace{-0.1em}
\subsection{Signal Model}
We consider a MU-MISO downlink system with secure transmission. 
Alice is equipped with an FCLA consisting of $M$ flexible circular arrays (FCAs), with each FCA equipped with $N$ antennas distributed along the circular track of radius $\rho$.
It is assumed that there are $K$ IRs, and AN is employed to interfere with the Eve, represented by $\boldsymbol{w}_e \sim \mathcal{CN}(0,\boldsymbol{R}_e)$.
We also assume that only the signal intended to IR 1 is confidential and should be kept secure from the Eve, while those intended to other IRs serve as interference at Eve.
The transmit signal intended for the IRs, together with the AN directed at Eve, is given by
\begin{align}
  \boldsymbol{x} = \boldsymbol{W}\boldsymbol{s}+\boldsymbol{w}_e,
\end{align}
where $\boldsymbol{W} = \left[\boldsymbol{w}_1, \dots, \boldsymbol{w}_K\right]\in\mathbb{C}^{MN\times K}$ denotes the transmit beamforming matrix.
$\boldsymbol{s} = \left[s_1, \dots, s_K\right]^T \in \mathbb{C}^{K\times 1}$ denotes the transmitted symbols sent to the IRs, with $\mathbb{E}\{\boldsymbol{s}\boldsymbol{s}^H\}=\boldsymbol{I}_K$.
The received signals at the $k$-th legitimate receiver, $k \in \mathcal{K}, \mathcal{K} \triangleq \{1,\dots,K\}$, and Eve are respectively given by
\begin{align}
  y_k
  = \boldsymbol{h}_k^H\boldsymbol{w}_k s_k
  + \boldsymbol{h}_k^H\sum_{k'\neq k}^K\boldsymbol{w}_{k'}s_{k'}
  + \boldsymbol{h}_k^H\boldsymbol{w}_e + n_k,
  \label{RxSigUser}
  \\
  y_e
  = \boldsymbol{h}_e^H\boldsymbol{w}_1 s_1
  + \boldsymbol{h}_e^H\sum_{k'\neq1}^K\boldsymbol{w}_{k'}s_{k'}
  + \boldsymbol{h}_e^H\boldsymbol{w}_e + n_e,
  \label{RxSigEve}
\end{align}
where $\boldsymbol{h}_k$ and $\boldsymbol{h}_e$ denote the channels from Alice to the $k$-th receiver and Eve, respectively. Moreover, $s_k$ denotes the data stream intended for IR $k$, while $n_k \sim \mathcal{CN}(0, \sigma_k^2)$ and $n_e \sim \mathcal{CN}(0, \sigma_e^2)$ denote the additive white Gaussian noise at the $k$-th IR and Eve, respectively.
Then, the signal-to-interference-plus-noise ratios (SINRs) at the $k$-th legitimate IR and Eve are given by, respectively,
\begin{align}
  {\rm \gamma}_k =
  \frac{\boldsymbol{h}_k^H\boldsymbol{w}_k\boldsymbol{w}_k^H\boldsymbol{h}_k }
  { \sum_{k'\neq k}^{K}\boldsymbol{h}_k^H\boldsymbol{w}_{k'}\boldsymbol{w}_{k'}^H\boldsymbol{h}_k + \boldsymbol{h}_k^H\boldsymbol{R}_{e}\boldsymbol{h}_k + \sigma_{k}^2 },\label{SINR_U}\\
  {\rm \gamma}_e =
  \frac{ \boldsymbol{h}_e^H\boldsymbol{w}_1\boldsymbol{w}_1^H\boldsymbol{h}_e }
  { \sum_{k'\neq 1}^{K}\boldsymbol{h}_e^H\boldsymbol{w}_{k'}\boldsymbol{w}_{k'}^H\boldsymbol{h}_e + \boldsymbol{h}_e^H\boldsymbol{R}_{e}\boldsymbol{h}_e + \sigma_{e}^2 }.\label{SINR_E}
\end{align}
\vspace{-2em}
\subsection{Channel Model}
Given that the signal propagation distance is significantly larger than the size of the moving region, the far-field response is adopted for channel modeling \cite{zlp2}.
Specifically, the path angles and complex gains for each link remain unchanged regardless of the antenna positions.
The geometric channel model is employed, assuming the same number of paths for all nodes \cite{zlp3,XIU3}.
Let $L_u$ denote the number of transmit paths from Alice to the $u$-th receiver, where $u \in \mathcal{U}$, $\mathcal{U}\triangleq\{1,\cdots,K,K+1\}$, and $u=K+1$ corresponds to Eve.
The elevation and azimuth angles of the $l$-th path of the $u$-th receiver's channel are denoted by $\theta_{u}^{l}$ and $\phi_{u}^{l}$, respectively.
In the polar coordinate system, the angle of the $n$-th element in the $m$-th FCA is denoted by $\varphi_{m,n} \in [0,2\pi]$, where $m \in \mathcal{M}, \mathcal{M}\triangleq\{1,\cdots,M\}$ and $n \in \mathcal{N}, \mathcal{N}\triangleq\{1,\cdots,N\}$.
The $m$-th FCA is capable of moving along the $z$-axis, with a height of $z_m \in \mathcal{Z}, \mathcal{Z}\triangleq [0,A]$.
Then, the $(m,n)$-th transmit MA position in the polar coordinate system is denoted by $\boldsymbol{t}_{m,n}=\left[\rho\cos\varphi_{m,n}, \rho\sin\varphi_{m,n}, z_m\right]^T$.
The signal propagation distance difference between the $(m,n)$-th transmitting MA position $\boldsymbol{t}_{m,n}$ and the reference point $\boldsymbol{o}^t$ is given by
\begin{align}
  \chi_{u}^{l}(\varphi_{m,n},z_m) & = \rho\cos\varphi_{m,n}\sin\theta_{u}^{l}\cos\phi_{u}^{l} +\rho\sin\varphi_{m,n}\notag\\ & \times\sin\theta_{u}^{l}\sin\phi_{u}^{l}   + z_m\cos\theta_{u}^{l}\triangleq \boldsymbol{t}_{m,n}^T\boldsymbol{\varrho}_{u}^{l},
\end{align}
where $\boldsymbol{\varrho}_{u}^{l} = \left[\varrho_{u}^{l,x}, \varrho_{u}^{l,y}, \varrho_{u}^{l,z}\right]^T$ represents the direction vector associated with the $l$-th path of receiver $u$, with components $\varrho_{u}^{l,x} = \sin\theta_{u}^{l}\cos\phi_{u}^{l}$, $\varrho_{u}^{l,y} = \sin\theta_{u}^{l}\sin\phi_{u}^{l}$, and $\varrho_{u}^{l,z} = \cos\theta_{u}^{l}$.
Therefore, the field response vector (FRV) at $\boldsymbol{t}_{m,n}$ is given by
\begin{align}
  &\tilde{\boldsymbol{g}}_{u}(\varphi_{m,n},z_m) = \notag\\
  &\left[
    e^{\jmath\frac{2\pi}{\lambda}\chi^{1}_{u}(\varphi_{m,n},z_m)},\cdots,
    e^{\jmath\frac{2\pi}{\lambda}\chi^{L_u}_{u}(\varphi_{m,n},z_m)}
  \right]
  ^T \in \mathbb{C}^{L_u\times 1},
\end{align}
where $\lambda$ is the wavelength of the carrier signal.
Considering the $m$-th FCA with $N$ antennas, the corresponding field response matrix (FRM) is
\begin{align}
  \boldsymbol{g}_{u}(\boldsymbol{\varphi}_{m}&,z_m)  = \notag\\
  &\left[\tilde{\boldsymbol{g}}_{u}(\boldsymbol{\varphi}_{m,1},z_m),\cdots,\tilde{\boldsymbol{g}}_{u}(\boldsymbol{\varphi}_{m,N},z_m)\right] \in \mathbb{C}^{L_u\times N},
\end{align}
where $\boldsymbol{\varphi}_{m} = [\varphi_{m,1},\cdots,\varphi_{m,N}]^T$ collects the element angles of the $m$-th FCA.
Consequently, the FRM of the link from Alice to receiver $u$ for all transmit MAs is given by
\begin{align}
  \boldsymbol{G}_u(\boldsymbol{\varphi},&\boldsymbol{z}) = \notag \\
  &\left[
    \boldsymbol{g}_{u}(\boldsymbol{\varphi}_{1},z_{1}), \cdots, \boldsymbol{g}_{u}(\boldsymbol{\varphi}_{M},z_{M})
  \right] \in \mathbb{C}^{L_u\times MN}.
\end{align}
Let $\boldsymbol{\Sigma}_u = \text{diag}\{\beta_{u}^{1},\cdots,\beta_{u}^{L_u}\} \in \mathbb{C}^{L_u\times L_u}$ be the path response matrix (PRM) from Alice to the $u$-th receiver. Then, the channel matrix at the $u$-th receiver is given by
\begin{align}
  \boldsymbol{h}_u^H(\boldsymbol{\varphi},\boldsymbol{z}) = \boldsymbol{1}^H\boldsymbol{\Sigma}_u\boldsymbol{G}_u(\boldsymbol{\varphi},\boldsymbol{z}) \in \mathbb{C}^{1\times MN},\forall u\in\mathcal{U}.
\end{align}
\vspace{-2.5em}
\subsection{Problem Formulation}
In this letter, we aim to maximize the achievable sum rate at all IRs by jointly optimizing the transmit beamforming, AN covariance, and antenna positions. The resulting optimization problem is formulated as
\begin{align}
  \underset{\boldsymbol{W},\boldsymbol{R}_e,\boldsymbol{\varphi},\boldsymbol{z}}{\max}\
  & \sum_{k=1}^{K}\log(1+\gamma_k)\label{p1}\\ {\rm s.t.}\
  & \log(1+\gamma_e) \leq \gamma_{th}^e,\tag{\ref{p1}a} \label{Const_1}\\
  & \mathrm{Tr}(\boldsymbol{W}^H\boldsymbol{W}) + \mathrm{Tr}(\boldsymbol{R}_e) \leq P,\tag{\ref{p1}b} \label{Const_2}\\
  & \varphi_{m,n+1}-\varphi_{m,n} \geq \varphi_{th}, \forall n\in\mathcal{N}\setminus\{N\},\notag\\
  & \varphi_{m,1}+2\pi-\varphi_{m,N} \geq \varphi_{th}, \forall m\in\mathcal{M},\tag{\ref{p1}c} \label{Const_3}\\
  & \varphi_{m,n}\in[0,2\pi],\forall m\in\mathcal{M}, \forall n\in\mathcal{N},\tag{\ref{p1}d} \label{Const_4}\\
  & z_{m+1}-z_m \geq z_{th}, \forall m\in\mathcal{M}\setminus\{M\},\tag{\ref{p1}e} \label{Const_5}\\
  & z_m\in \mathcal{Z},\forall m\in\mathcal{M},\tag{\ref{p1}f} \label{Const_6}
\end{align}
where $P$ is the total transmit power, $\varphi_{th}$ is the minimal angle between adjacent elements within one FCA, and $z_{th}$ is the minimal spacing between adjacent FCAs. $\varphi_{th}$ and $z_{th}$ are designed to ensure that the antenna elements are not positioned too closely together, thereby avoiding severe mutual coupling. $\gamma_{th}^e$ denotes the maximum allowable rate for Eve to ensure secure transmission. Note that Problem (\ref{p1}) is intractable due to the highly non-concave objective function and the coupling among the optimization variables.

\section{Proposed BCD Algorithm}
To solve (\ref{p1}), we first reformulate the objective function in (\ref{p1}) using the Lagrangian dual transform and then solve the resulting problem via a BCD-based framework that combines PGD, SDR, and successive convex approximation (SCA) methods.
\vspace{-1.5em}
\subsection{Problem Reformulation}
Utilizing the Lagrangian dual transformation method, we transform the sum-of-logarithms objective in (\ref{p1}) into an equivalent sum-of-ratios form, i.e.,
\begin{small}
\begin{align}
  &\mathcal{I}_a(\boldsymbol{W},\boldsymbol{R}_e,\boldsymbol{\varphi},\boldsymbol{z},\boldsymbol{\eta}) =  \sum_{k=1}^{K} \log(1+\eta_k) - \eta_k \notag\\
  & + \underbrace{\frac{(1+\eta_k)\vert\boldsymbol{h}_k^H(\boldsymbol{\varphi},\boldsymbol{z})\boldsymbol{w}_k\vert^2}
  {\sum_{k'=1}^{K}\vert\boldsymbol{h}_{k}^H(\boldsymbol{\varphi},\boldsymbol{z})\boldsymbol{w}_{k'}\vert^2 + \boldsymbol{h}_k^H(\boldsymbol{\varphi},\boldsymbol{z})\boldsymbol{R}_e\boldsymbol{h}_k(\boldsymbol{\varphi},\boldsymbol{z}) + \sigma^2_k}}_{\mathcal{I}_b(\boldsymbol{W},\boldsymbol{R}_e,\boldsymbol{\varphi},\boldsymbol{z})},
\end{align}
\end{small}
where $\boldsymbol{\eta} = [\eta_1,\cdots,\eta_K]^T \in \mathbb{R}^{K\times 1}$ is the slack variable. Since $\mathcal{I}_a$ is concave with respect to (w.r.t.) $\eta_k$, the optimal $\eta_k^\star$ is obtained from the first-order optimality condition as $\eta_k^\star = \gamma_k$.
It is readily seen that only the last term of $\mathcal{I}_a(\boldsymbol{W},\boldsymbol{R}_e,\boldsymbol{\varphi},\boldsymbol{z},\boldsymbol{\eta})$ is involved in the optimization on $\boldsymbol{W},\boldsymbol{R}_e,\boldsymbol{\varphi},\boldsymbol{z}$ with $\boldsymbol{\eta}$ fixed.
To proceed further, we first define an auxiliary variable $\boldsymbol{\varpi} \triangleq [\varpi_1,\cdots,\varpi_K]^T \in \mathbb{C}^{K\times 1}$, and then by applying a quadratic transform, we further recast the last term as
\vspace{-0.6em}
\begin{small}
\begin{align}\label{I_b}
  \mathcal{I}_b(\boldsymbol{W},\boldsymbol{R}_e,\boldsymbol{\varphi},\boldsymbol{z},&\boldsymbol{\varpi})
  = \sum_{k=1}^{K} \{2(1+\eta_k)\varpi_k\sqrt{\vert\boldsymbol{h}_k^H(\boldsymbol{\varphi},\boldsymbol{z})\boldsymbol{w}_k\vert^2}\notag\\[-0.4em]
  & -(1+\eta_k)\vert\varpi_k\vert^2(\sum_{k' = 1}^K\vert\boldsymbol{h}_k^H(\boldsymbol{\varphi},\boldsymbol{z})\boldsymbol{w}_{k'}\vert^2  \notag\\
  & + \boldsymbol{h}_k^H(\boldsymbol{\varphi},\boldsymbol{z})\boldsymbol{R}_e\boldsymbol{h}_k(\boldsymbol{\varphi},\boldsymbol{z}) + \sigma^2_k)\} + \text{const},
\end{align}
\end{small}%
where const denotes a constant term independent of the optimization variables.
It is observed that $\mathcal{I}_b(\boldsymbol{W},\boldsymbol{R}_e,\boldsymbol{\varphi},\boldsymbol{z},\boldsymbol{\varpi})$ is still non-convex owing to the strong coupling between optimization variables.
To address this, we propose a BCD-based framework to obtain a high-quality solution.
\vspace{-1.5em}
\subsection{Updating Auxiliary Variables}
With $\boldsymbol{W}$, $\boldsymbol{R}_e$, $\boldsymbol{\varphi}$, and $\boldsymbol{z}$ fixed, $\mathcal{I}_b$ is concave with respect to $\varpi_k$ for all $k$. Hence, the optimal closed-form update of $\varpi_k$ is given by
\begin{small}
\begin{align}
  \varpi_k^\star = 
  \frac{\sqrt{\vert\boldsymbol{h}_k^H(\boldsymbol{\varphi},\boldsymbol{z})\boldsymbol{w}_k\vert^2}}{\sum_{k' = 1}^K\vert\boldsymbol{h}_k^H(\boldsymbol{\varphi},\boldsymbol{z})\boldsymbol{w}_{k'}\vert^2 + \boldsymbol{h}_k^H(\boldsymbol{\varphi},\boldsymbol{z})\boldsymbol{R}_e\boldsymbol{h}_k(\boldsymbol{\varphi},\boldsymbol{z}) + \sigma^2_k}, \forall k,
\end{align}
\end{small}
\vspace{-2.5em}
\subsection{Updating Transmit Beamforming}
Given $\boldsymbol{\varpi}$, $\boldsymbol{\varphi}$, and $\boldsymbol{z}$, we focus on the optimization of $\boldsymbol{W}$ and $\boldsymbol{R}_e$. The non-convexity of Problem (\ref{p1}) mainly arises from the quadratic terms with respect to $\boldsymbol{w}_k$ in (\ref{Const_1}) and (\ref{I_b}). To address this issue, the SDR method is employed. Specifically, we first introduce auxiliary matrices $\{\boldsymbol{R}_k\}_{k=1}^{K}$ with $\boldsymbol{R}_k = \boldsymbol{w}_k\boldsymbol{w}_k^H$, where each $\boldsymbol{R}_k$ is rank-one and positive semidefinite.
We define $\boldsymbol{R} = \sum_{k = 1}^{K}\boldsymbol{R}_k + \boldsymbol{R}_e$.
By incorporating (\ref{I_b}) into (\ref{p1}) and leveraging the monotonic property of the logarithm function, we recast the problem in (\ref{p1}) as
\begin{align}
  \underset{\{\boldsymbol{R}_{k}\}^{K}_{k=1},\boldsymbol{R}}{\max}\
  & \sum_{k=1}^{K} \{2(1+\eta_k)\varpi_k\sqrt{\boldsymbol{h}_k^H(\boldsymbol{\varphi},\boldsymbol{z})\boldsymbol{R}_k\boldsymbol{h}_k(\boldsymbol{\varphi},\boldsymbol{z})}\notag\\
  & -(1+\eta_k)\vert\varpi_k\vert^2(\boldsymbol{h}_k^H(\boldsymbol{\varphi},\boldsymbol{z})\boldsymbol{R}\boldsymbol{h}_k(\boldsymbol{\varphi},\boldsymbol{z}) + \sigma^2_k)\}\label{p2}\\
  {\mathrm{s.t.}}\
  &
  \frac{\boldsymbol{h}_e^H\boldsymbol{R}_1\boldsymbol{h}_e}
  {\boldsymbol{h}_e^H\{\boldsymbol{R}-\boldsymbol{R}_1\}\boldsymbol{h}_e + \sigma^2_e}
  \leq \Gamma_{th}^e, \label{p2a} \tag{\ref{p2}a}\\
  &
  \boldsymbol{R}-\boldsymbol{R}_1 \succeq 0,\mathrm{Tr}(\boldsymbol{R}) \leq P,\label{p2b} \tag{\ref{p2}b} \\
  &
  \boldsymbol{R}_k \succeq 0, \mathrm{rank}(\boldsymbol{R}_k) = 1, \forall k.\label{p2c} \tag{\ref{p2}c}
\end{align}
where $\Gamma_{th}^e = 2^{\gamma_{th}^e}-1$.
By dropping the rank-one constraint in (\ref{p2}c), Problem (\ref{p2}) becomes a standard semidefinite programming problem and can be solved via the CVX tool.
Denote the optimal solutions of the relaxed problem by $\{\tilde{\boldsymbol{R}}_k\}_{k=1}^{K}$ and $\tilde{\boldsymbol{R}}$.
It is worth noting that if the obtained $\{\tilde{\boldsymbol{R}}_k\}_{k=1}^{K}$ satisfies the rank-one condition, it directly serves as the optimal solution to the original rank-constrained beamforming subproblem in (\ref{p2}).
While the relaxation is not always tight, we can always construct an optimal rank-one solution. Specifically, 
we can retrieve the optimal $\boldsymbol{R}^\star$ and rank-one $\{\boldsymbol{R}_k^\star\}_{k=1}^{K}$ via
\begin{align} \label{SDR_optimal}
  &\boldsymbol{R}^\star = \tilde{\boldsymbol{R}},\boldsymbol{w}_k^\star = (\boldsymbol{h}_k^H(\boldsymbol{\varphi},\boldsymbol{z})\tilde{\boldsymbol{R}}_k\boldsymbol{h}_k(\boldsymbol{\varphi},\boldsymbol{z}))^{-1/2}\tilde{\boldsymbol{R}}_k\boldsymbol{h}_k(\boldsymbol{\varphi},\boldsymbol{z}), \notag\\
  &\boldsymbol{R}_k^\star = \boldsymbol{w}_k^\star\boldsymbol{w}_k^{\star H}, \boldsymbol{R}_e^\star = \boldsymbol{R}^\star - \sum_{k=1}^{K}\boldsymbol{R}_k^\star.
\end{align}

The proof is provided in Appendix \ref{AppSDR}. The complexity for updating $\boldsymbol{W}$ and $\boldsymbol{R}_e$ is $\mathcal{O}(M^{6.5}N^{6.5}K^{3.5})$.

\subsection{Updating Transmit Antenna Placement}
In this subsection, we update the transmit antenna placement, including the rotation angles $\boldsymbol{\varphi}$ and vertical coordinates $\boldsymbol{z}$, while fixing $\boldsymbol{\varpi}$, $\boldsymbol{W}$, and $\boldsymbol{R}_e$. The non-convexity of Problem (\ref{p1}) mainly arises from the transformed objective function $\mathcal{I}_b$ in (\ref{I_b}) and the constraints in (\ref{Const_1}) and (\ref{Const_3})-(\ref{Const_6}). To handle this issue, we reformulate the problem as
\begin{align}
  \underset{\boldsymbol{\varphi},\boldsymbol{z}}{\max}\
  & \sum_{k=1}^{K} 2(1+\eta_k)\varpi_k\Re\{\boldsymbol{h}_k^H(\boldsymbol{\varphi},\boldsymbol{z})\boldsymbol{w}_k\}\notag\\&
  -(1+\eta_k)\varpi_k^2\left(\boldsymbol{h}_{k}^H(\boldsymbol{\varphi},\boldsymbol{z})\boldsymbol{\Xi}_s\boldsymbol{h}_{k}(\boldsymbol{\varphi},\boldsymbol{z}) + \sigma^2_k\right) \label{p3} \\
  {\rm s.t.}\
  & {\boldsymbol{h}_e^H(\boldsymbol{\varphi},\boldsymbol{z})\boldsymbol{\Xi}_e\boldsymbol{h}_e(\boldsymbol{\varphi},\boldsymbol{z})}
  \geq -\Gamma_{th}^e\sigma_e^2
  \tag{\ref{p3}a} \label{p3a}\\
  & (\text{\ref{Const_3}})-(\text{\ref{Const_6}})\notag
\end{align}
where $\boldsymbol{\Xi}_e = \Gamma_{th}^e\left(\sum_{k'\neq 1}^{K}\boldsymbol{w}_{k'}\boldsymbol{w}_{k'}^H + \boldsymbol{R}_e\right) -\boldsymbol{w}_1\boldsymbol{w}_1^H$ and $\boldsymbol{\Xi}_s = \sum_{k'=1}^{K}\boldsymbol{w}_{k'}\boldsymbol{w}_{k'}^H+\boldsymbol{R}_e$.
We employ a BCD-based algorithm to alternately update the rotation angles $\boldsymbol{\varphi}$ and vertical coordinates $\boldsymbol{z}$. To this end, we adopt PGD with Nesterov acceleration to speed up convergence.
Let $\boldsymbol{\xi} \in \{\boldsymbol{\varphi}, \boldsymbol{z}\}$ denote the optimization variable being updated in the BCD iteration, and $\nabla \mathcal{I}_b(\boldsymbol{\xi}) \in \{\nabla \mathcal{I}_b(\boldsymbol{\varphi}), \nabla \mathcal{I}_b(\boldsymbol{z})\}$ denote the gradient vector of the objective function with respect to $\boldsymbol{\xi}$.
The antenna position variables are updated iteratively as
\begin{align}
  &\text{(Step 1)}\ \boldsymbol{\zeta}^{(t+1)} = \boldsymbol{\iota}^{(t)} + \tau^{(t)} \nabla \mathcal{I}_b(\boldsymbol{\iota}^{(t)}), \\
  &\text{(Step 2)}\ \boldsymbol{\xi}^{(t+1)} = \arg \min \Vert \boldsymbol{\xi} - \boldsymbol{\zeta}^{(t+1)} \Vert \notag\\
  & ~~~~~~~~~~~~~~~~~~~~~~ \text{s.t.}\  (\text{\ref{p3}a}), (\text{\ref{Const_3}})-(\text{\ref{Const_6}}) \\
  &\text{(Step 3)}\ \boldsymbol{\iota}^{(t+1)} = \boldsymbol{\xi}^{(t+1)} + \alpha_{t+1}(\boldsymbol{\xi}^{(t+1)} - \boldsymbol{\xi}^{(t)}),
\end{align}
where $\boldsymbol{\zeta}^{(t+1)}$ is an intermediate variable and the superscript $t$ denotes the iteration index. In Step 1, $\tau^{(t)}\geq 0$ is determined by backtracking line search. In Step 2, the projection operation ensures that the updated antenna positions satisfy (\ref{p3}a) and (\ref{p1}c)-(\ref{p1}f). In Step 3, Nesterov acceleration is applied with $\alpha_{t+1}=\frac{q_{t+1}-1}{q_{t+1}}$ and $q_{t+1}=\frac{1+\sqrt{1+4q_t^2}}{2}$, where $q_0=1$.
Let $\tilde{\mathcal{F}}_{u,v}(\boldsymbol{\varphi},\boldsymbol{z}) = \boldsymbol{h}_u^H(\boldsymbol{\varphi},\boldsymbol{z})\boldsymbol{\Xi}_v\boldsymbol{h}_u(\boldsymbol{\varphi},\boldsymbol{z})$ with $v\in\{s,e\}$ and $\mathcal{\tilde L}_{k}(\boldsymbol{\varphi},\boldsymbol{z})=\Re\{\boldsymbol{h}_k^H(\boldsymbol{\varphi},\boldsymbol{z})\boldsymbol{w}_k\}$. Then, the gradient vector of the objective function $\mathcal{I}_b(\boldsymbol{\xi})$ at $\boldsymbol{\iota}^{(t)}$ can be written as
\begin{small}
\begin{align}
  \nabla \mathcal{I}_b(\boldsymbol{\xi})
  = \sum_{k=1}^{K}(1+\eta_k)\varpi_k\left(2\nabla_{\boldsymbol{\xi}}\mathcal{\tilde{L}}_{k}(\boldsymbol{\xi})
  -\varpi_k\nabla_{\boldsymbol{\xi}}\mathcal{\tilde{F}}_{k,s}(\boldsymbol{\xi})\right),
\end{align}
\end{small}%
where $\nabla_{\boldsymbol{\xi}}\mathcal{\tilde{L}}_{k}(\boldsymbol{\xi})$ and $\nabla_{\boldsymbol{\xi}}\mathcal{\tilde{F}}_{k,s}(\boldsymbol{\xi})$ denote the gradients of $\mathcal{\tilde{L}}_{k}(\boldsymbol{\xi})$ and $\mathcal{\tilde{F}}_{k,s}(\boldsymbol{\xi})$ at $\boldsymbol{\xi}$.
If $\boldsymbol{\xi}=\boldsymbol{\varphi}$, the entries of $\nabla_{\boldsymbol{\varphi}}\tilde{\mathcal{F}}_{u,v}$ and $\nabla_{\boldsymbol{\varphi}}\tilde{\mathcal{L}}_{u}$ are $\frac{\partial\tilde{\mathcal{F}}_{u,v}}{\rho\partial\varphi_{m,n}}$ and $\frac{\partial\tilde{\mathcal{L}}_{u}}{\rho\partial\varphi_{m,n}}$. If $\boldsymbol{\xi}=\boldsymbol{z}$, the entries of $\nabla_{\boldsymbol{z}}\tilde{\mathcal{F}}_{u,v}$ and $\nabla_{\boldsymbol{z}}\tilde{\mathcal{L}}_{u}$ are $\frac{\partial\tilde{\mathcal{F}}_{u,v}}{\partial z_m}$ and $\frac{\partial\tilde{\mathcal{L}}_{u}}{\partial z_m}$.
We expand $\tilde{\mathcal{F}}_{u,v}(\boldsymbol{\varphi},\boldsymbol{z})$ as
\begin{align}
  &\tilde{\mathcal{F}}_{u,v}(\boldsymbol{\varphi},\boldsymbol{z})
  =
  \sum_{i=1}^{N_t}
  \sum_{l=1}^{L}
  \vert\beta_{u}^{l}\vert^2
  \Xi_{v}^{i,i}
  +
  \sum_{i=1}^{N_t}
  \sum_{l=1}^{L-1}
  \sum_{p=l+1}^{L}
  2\mu_{u,v}^{i,i,l,p}
  \notag\\ &
  \times\cos{\kappa_{i,i,l,p}^{u,v}}
  +
  \sum_{i=1}^{N_t-1}
  \sum_{j=i+1}^{N_t}
  \sum_{l=1}^{L}
  \sum_{p=1}^{L}
  2\mu_{u,v}^{i,j,l,p}
  \cos{\kappa_{i,j,l,p}^{u,v}},
\end{align}
where $N_t = MN$ is the total number of antennas, $\mu_{u,v}^{i,j,l,p}=\vert\beta_{u}^{l}\vert\vert\beta_{u}^{p}\vert\vert\Xi_{v}^{i,j}\vert$, and $\kappa_{u,v}^{i,j,l,p}=\angle\beta_{u}^{l} - \angle\beta_{u}^{p}+\angle\Xi_{v}^{i,j}+\frac{2\pi}{\lambda}(\chi_{u}^{l}(\varphi_{i}, z_{m_i})-\chi_{u}^{p}(\varphi_{j}, z_{m_j }))$, where $i = (m-1)N+ n$ and $m_i = \lceil \frac{i}{N} \rceil$. We expand $\mathcal{\tilde L}_{u}(\boldsymbol{\varphi},\boldsymbol{z})$ as
\begin{align}
  \mathcal{\tilde{L}}_u(\boldsymbol{\varphi}, \boldsymbol{z})
  =
  \sum_{i=1}^{N_t}
  \sum_{l=1}^{L}
  \hat{\mu}_{u}^{i,l}
  \cos(
  \hat\kappa_{u}^{i,l}
  ),
\end{align}
where $\hat{\mu}_{u}^{i,l} =  \vert \beta_{u}^{l} \vert \vert w_{i} \vert$ and $\hat\kappa_{u}^{i,l} = \frac{2\pi}{\lambda}\chi_{u}^{l}(\varphi_{i}, z_{m_i})+\angle{\beta_{u}^{l}} +\angle{w_i}$.

The problem is intractable due to the constraints in (\ref{p3a}). To deal with this issue, the SCA method is applied. By exploiting the second-order Taylor expansion, a quadratic lower bound of the non-convex term is constructed as
\begin{align}\label{p3a_expansion}
  \tilde{\mathcal{F}}_{e,e}(\boldsymbol{\xi}) \geq \tilde{\mathcal{F}}_{e,e}^{lb,r}(\boldsymbol{\xi}) \triangleq &
  \tilde{\mathcal{F}}_{e,e}(\boldsymbol{\xi}^{(r)}) + \nabla_{\boldsymbol{\xi}}\tilde{\mathcal{F}}_{e,e}(\boldsymbol{\xi}^{(r)})^T(\boldsymbol{\xi}-\boldsymbol{\xi}^{(r)}) \notag\\ &
  - \frac{\delta_e^{(r)}}{2}(\boldsymbol{\xi}-\boldsymbol{\xi}^{(r)})^T(\boldsymbol{\xi}-\boldsymbol{\xi}^{(r)})
\end{align}
where the positive real number $\delta_e^{(r)}$ satisfies $\delta_e^{(r)}\boldsymbol{I}_{N_t} \succeq \nabla^2_{\boldsymbol{\xi}}\tilde{\mathcal{F}}_{e,e}(\boldsymbol{\xi})$, and $\nabla^2_{\boldsymbol{\xi}}\tilde{\mathcal{F}}_{e,e}(\boldsymbol{\xi})$ denotes the Hessian matrix of $\tilde{\mathcal{F}}_{e,e}(\boldsymbol{\xi})$ w.r.t.\ $\boldsymbol{\xi}$.
For $\boldsymbol{\xi}=\boldsymbol{\varphi}$ and $\boldsymbol{\xi}=\boldsymbol{z}$, the Hessian matrices $\nabla^2_{\boldsymbol{\varphi}}\tilde{\mathcal{F}}_{e,e}(\boldsymbol{\varphi})$ and $\nabla^2_{\boldsymbol{z}}\tilde{\mathcal{F}}_{e,e}(\boldsymbol{z})$ are given by, respectively,
\begin{small}
\begin{align}
  \nabla^2_{\boldsymbol{\varphi}}\tilde{\mathcal{F}}_{e,e}(\boldsymbol{\xi})
  &=
  \left[
  \frac{\partial^2\tilde{\mathcal{F}}_{e,e}(\boldsymbol{\xi})}
  {\rho^2 \partial \varphi_i \partial \varphi_j}
  \right]_{i,j=1}^{N_t}, \notag\\
  \nabla^2_{\boldsymbol{z}}\tilde{\mathcal{F}}_{e,e}(\boldsymbol{\xi})
  &=
  \left[
  \frac{\partial^2\tilde{\mathcal{F}}_{e,e}(\boldsymbol{\xi})}
  {\partial z_m \partial z_n}
  \right]_{m,n=1}^{M}.
\end{align}
\end{small}%
Then, the resulting convex subproblem can be written as
\begin{align}\label{p4}
  \min_{\boldsymbol{\xi}} \Vert\boldsymbol{\xi}-\boldsymbol{\zeta}^{(r)}\Vert_2^2 
  ~\text{s.t.} 
  \begin{cases} 
  \displaystyle \tilde{\mathcal{F}}_{e,e}^{lb,r}(\boldsymbol{\xi}) \geq -\Gamma_{th}^e\sigma_e^2,\\ 
  \displaystyle \left(\text{\ref{Const_3}}\right)-\left(\text{\ref{Const_4}}\right),\text{if}~ \boldsymbol{\xi}=\boldsymbol{\varphi},\\
  \displaystyle \left(\text{\ref{Const_5}}\right)-\left(\text{\ref{Const_6}}\right),\text{if}~ \boldsymbol{\xi}=\boldsymbol{z},
  \end{cases}
\end{align}
which can be solved using the CVX tool. The complexities for updating $\boldsymbol{\varphi}$ and $\boldsymbol{z}$ are $\mathcal{O}(M^{3.5}N^{3.5})$ and $\mathcal{O}(M^{3.5}N^{2})$, respectively.

\section{Simulation Results}
\vspace{-0.4em}
\begin{figure*}[t]
	\vspace{-2em}
	\setlength{\abovecaptionskip}{0.1cm}
	\centering
	\subfloat[Convergence of the proposed algorithm.\label{iteration}]{\includegraphics[width= 6 cm]{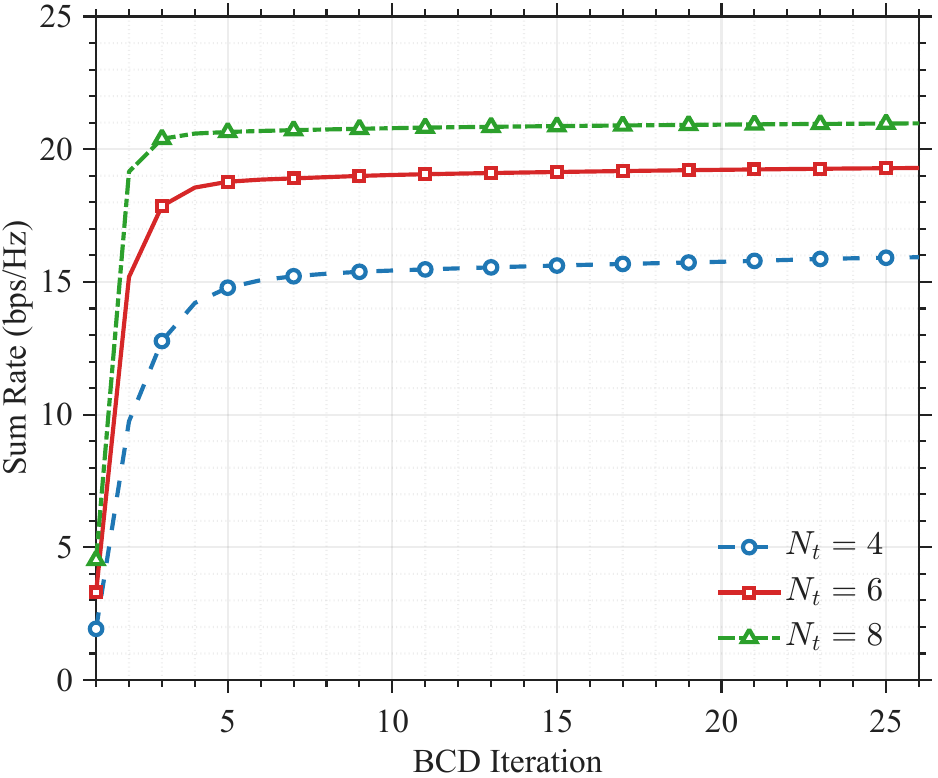}}%
	\hfil
	\subfloat[Sum rate versus transmit power.\label{PP}]{\includegraphics[width= 6 cm]{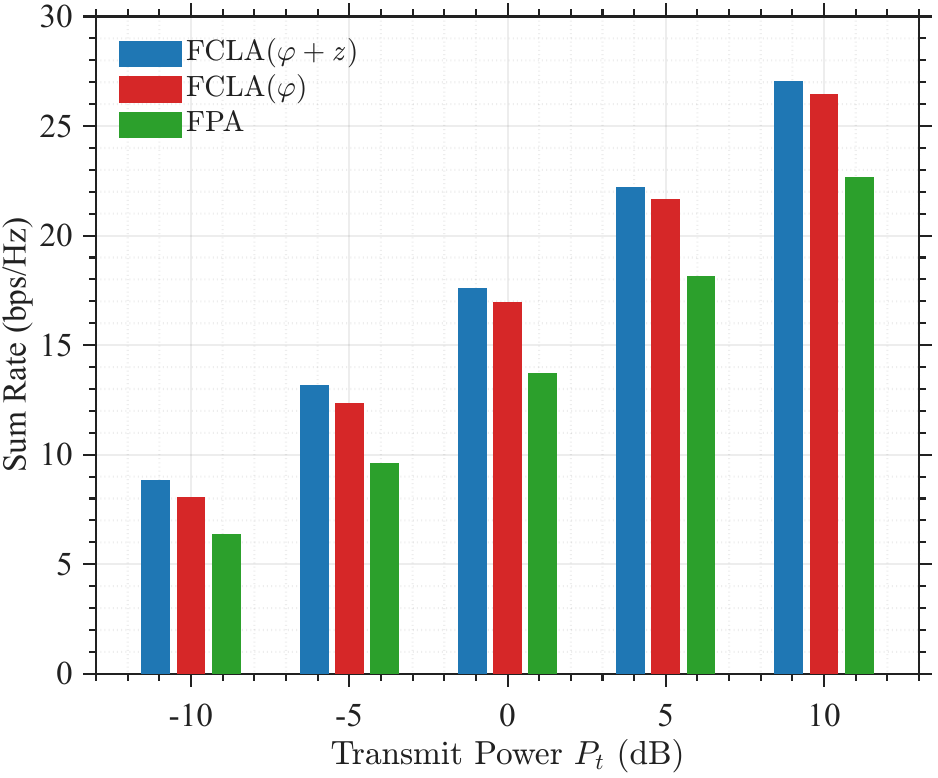}}
	\hfil
	\subfloat[Sum rate versus movable region size.\label{P_SINR}]{\includegraphics[width= 6 cm]{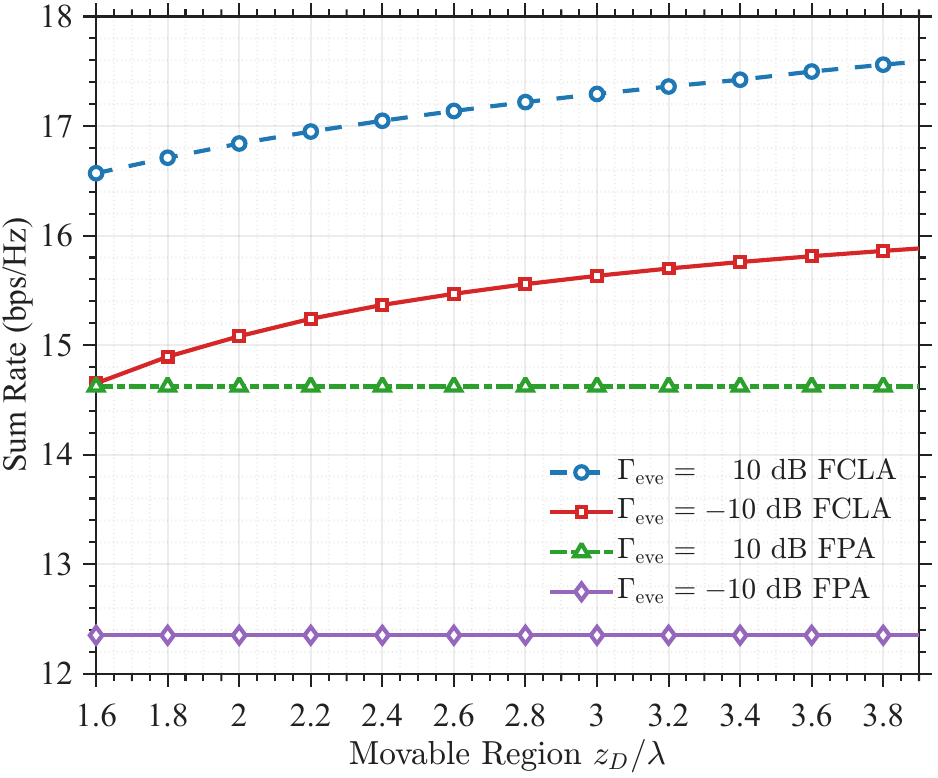}}%
	\hfil
	\caption{Performance comparison of different schemes.}
	\label{simu}
	\vspace{-2em}
\end{figure*}

In this section, we present numerical results to evaluate the performance of our proposed scheme with FCLA and compare it with that of the following 
three benchmark schemes: 1) \textbf{FPA}: Alice employs $N_t$ fixed-position antennas (FPAs) with adjacent spacing $\frac{\lambda}{2}$. 2) \textbf{FCLA($\boldsymbol{\varphi}$)}: Alice employs an FCLA with $N_t$ antennas, where only the angular variables $\boldsymbol{\varphi}$ are optimized while the vertical coordinates $\boldsymbol{z}$ are fixed. 3) \textbf{FCLA($\boldsymbol{\varphi},\boldsymbol{z}$)}: Alice employs an FCLA with joint optimization of $\boldsymbol{\varphi}$ and $\boldsymbol{z}$, which corresponds to the proposed scheme.

In our simulations, Alice is located at $(0,0)$ m, and the users are randomly distributed within a disk centered at $(40,0)$ m with radius $10$ m. Alice employs an FCLA with $M=3$ FCAs and $N=2$ antennas per FCA to serve $K=3$ users. Unless otherwise specified, each link has the same number of paths, i.e., $L_u=L=4$, and the path response matrix is given by $\boldsymbol{\Sigma}_u=\text{diag}\{\beta_u^1,\cdots,\beta_u^L\}$ with $\beta_u^l \sim \mathcal{CN}(0,\frac{c_u^2}{L})$, where $c_u^2=C_0 d_u^{-\alpha}$, $C_0=-30$ dB, and $\alpha=2.3$. The eavesdropping channel is generated in the same manner as the channels to the legitimate IRs. The elevation and azimuth angles $\theta_u^l$ and $\phi_u^l$ are independently drawn from the uniform distributions over $[\frac{\pi}{6},\frac{5\pi}{6}]$ and $[0,2\pi]$, respectively. The minimum vertical spacing between adjacent FCAs is set as $z_{th}=\frac{\lambda}{2}$ with $\lambda=0.1$ m. Moreover, the noise powers at the users and Eve are set to $\sigma_k^2=\sigma_e^2=-90$ dBm, the transmit power budget is $P=3$ dBW, and the secrecy threshold is set to $\Gamma_{th}^e=-10$ dB.

Fig. \ref{simu}(a) illustrates the convergence behavior of the proposed BCD algorithm under different system configurations. It is observed that the objective value increases rapidly and converges within about five iterations for all considered $N_t$. Moreover, the converged sum rate increases with $N_t$, since more antennas provide additional spatial degrees of freedom for joint beamforming, AN covariance matrix, and antenna-position optimization.

Fig. \ref{simu}(b) shows the achievable sum rate versus the transmit power $P_t$. As expected, the sum rates of all schemes increase with $P_t$. The proposed FCLA($\boldsymbol{\varphi},\boldsymbol{z}$) scheme consistently achieves the best performance, followed by FCLA($\boldsymbol{\varphi}$), while FPA performs the worst. This verifies that flexible cylindrical-array geometry brings significant gains over fixed arrays, and that jointly optimizing the angular and vertical antenna positions further improves the sum-rate performance.

Fig. \ref{simu}(c) depicts the achievable sum rate versus the movable region size $z_D/\lambda$ under different secrecy thresholds $\Gamma_{th}^e$. The sum rate of the FCLA-based schemes improves as $z_D/\lambda$ increases, because a larger movable region provides more flexibility for geometry adaptation. By contrast, the FPA performance remains unchanged since fixed arrays cannot exploit this additional spatial freedom. In addition, a looser secrecy constraint yields a higher achievable sum rate for both FCLA and FPA schemes.
\vspace{-1.1em}
\section{Conclusions}\label{Con}
In this letter, we have investigated an AN-aided secure MU-MISO system enhanced by a FCLA to maximize the system sum rate via joint optimization of transmit beamforming, AN covariance, and antenna positions under secrecy, power, and placement constraints. To solve it, we developed a BCD-based framework combining the Lagrangian dual transform, SDR, PGD and SCA methods. Numerical results showed fast convergence and substantial sum-rate gains of our proposed scheme over benchmark schemes. 
\vspace{-1em}
\begin{appendices}
  \section{Proof of Optimality of (\ref{SDR_optimal}) for Problem (\ref{p2})}\label{AppSDR}
  Let $\tilde{\boldsymbol{R}}_k$ denote the optimal solution obtained from the relaxed SDR problem. The constructed rank-one solution is given by $\boldsymbol{w}_k^\star = (\boldsymbol{h}_k^H\tilde{\boldsymbol{R}}_k\boldsymbol{h}_k)^{-1/2}\tilde{\boldsymbol{R}}_k\boldsymbol{h}_k$.
  First, if $\boldsymbol{R}_k^\star$ is a rank-one matrix, then we have
  \begin{align}
    \boldsymbol{h}_k^H\boldsymbol{R}_k^\star\boldsymbol{h}_k = \boldsymbol{h}_k^H\boldsymbol{w}_k^\star\boldsymbol{w}_k^{\star H}\boldsymbol{h}_k = \boldsymbol{h}_k^H\tilde{\boldsymbol{R}}_k\boldsymbol{h}_k.
  \end{align}
  Thus, the value of the objective function component associated with the desired signal power remains unchanged. Next, we show that $\tilde{\boldsymbol{R}}_k - \boldsymbol{R}_k^\star \succeq 0$. For any $\boldsymbol{v} \in \mathbb{C}^{N_t \times 1}$, it holds that
  \begin{small}
  \begin{align}
    \boldsymbol{v}^H(\tilde{\boldsymbol{R}}_k - \boldsymbol{R}_k^\star)\boldsymbol{v} &= \boldsymbol{v}^H\tilde{\boldsymbol{R}}_k\boldsymbol{v} - \boldsymbol{v}^H \boldsymbol{w}_k^\star\boldsymbol{w}_k^{\star H} \boldsymbol{v} \notag \\
    &= \boldsymbol{v}^H\tilde{\boldsymbol{R}}_k\boldsymbol{v} - (\boldsymbol{h}_k^H\tilde{\boldsymbol{R}}_k\boldsymbol{h}_k)^{-1} |\boldsymbol{v}^H\tilde{\boldsymbol{R}}_k\boldsymbol{h}_k|^2.
  \end{align}
\end{small}
  According to the Cauchy-Schwarz inequality, we have
  \begin{equation}
    (\boldsymbol{h}_k^H\tilde{\boldsymbol{R}}_k\boldsymbol{h}_k)(\boldsymbol{v}^H\tilde{\boldsymbol{R}}_k\boldsymbol{v}) \geq |\boldsymbol{v}^H\tilde{\boldsymbol{R}}_k\boldsymbol{h}_k|^2.
  \end{equation}
  It follows that $\boldsymbol{v}^H(\tilde{\boldsymbol{R}}_k - \boldsymbol{R}_k^\star)\boldsymbol{v} \geq 0$ holds true for any $\boldsymbol{v} \in \mathbb{C}^{N_t \times 1}$, i.e., $\tilde{\boldsymbol{R}}_k - \boldsymbol{R}_k^\star \succeq 0$. We can conclude that $\boldsymbol{R}_e^\star = \tilde{\boldsymbol{R}} - \sum_{k=1}^K \boldsymbol{R}_k^\star \succeq \tilde{\boldsymbol{R}} - \sum_{k=1}^K \tilde{\boldsymbol{R}}_k = \tilde{\boldsymbol{R}}_e \succeq 0$.
  Furthermore, to verify constraint (\ref{p2a}), we rewrite it as $\boldsymbol{h}_e^H\boldsymbol{R}_1\boldsymbol{h}_e \leq \frac{\Gamma_{th}^e}{1+\Gamma_{th}^e}(\boldsymbol{h}_e^H\boldsymbol{R}\boldsymbol{h}_e + \sigma^2_e)$. Recall that $\tilde{\boldsymbol{R}}_1 - \boldsymbol{R}_1^\star \succeq 0$ implies $\boldsymbol{h}_e^H\boldsymbol{R}_1^\star\boldsymbol{h}_e \leq \boldsymbol{h}_e^H\tilde{\boldsymbol{R}}_1\boldsymbol{h}_e$, and $\boldsymbol{R}^\star = \tilde{\boldsymbol{R}}$.
  Thus, we have
  \begin{align}
    \boldsymbol{h}_e^H\boldsymbol{R}_1^\star\boldsymbol{h}_e \leq \boldsymbol{h}_e^H\tilde{\boldsymbol{R}}_1\boldsymbol{h}_e &\leq \frac{\Gamma_{th}^e}{1+\Gamma_{th}^e}(\boldsymbol{h}_e^H\tilde{\boldsymbol{R}}\boldsymbol{h}_e + \sigma^2_e) \notag \\ &= \frac{\Gamma_{th}^e}{1+\Gamma_{th}^e}(\boldsymbol{h}_e^H\boldsymbol{R}^\star\boldsymbol{h}_e + \sigma^2_e),
  \end{align}
  which confirms that (\ref{p2a}) is satisfied.
  Consequently, all the constraints in the original problem are met.
  Based on the above, we can verify that $\{\boldsymbol{R}^\star, \{\boldsymbol{R}_k^\star\}_{k=1}^K, \boldsymbol{R}_e^\star\}$ is a feasible solution to the original rank-constrained beamforming subproblem in (\ref{p2}). Moreover, it attains the same objective value as the relaxed SDR solution and is therefore globally optimal for Problem (\ref{p2}).
\end{appendices}
\bibliographystyle{IEEEtran}
\bibliography{reference.bib}

\begin{thebibliography}{10}
\providecommand{\url}[1]{#1}
\csname url@samestyle\endcsname
\providecommand{\newblock}{\relax}
\providecommand{\bibinfo}[2]{#2}
\providecommand{\BIBentrySTDinterwordspacing}{\spaceskip=0pt\relax}
\providecommand{\BIBentryALTinterwordstretchfactor}{4}
\providecommand{\BIBentryALTinterwordspacing}{\spaceskip=\fontdimen2\font plus
\BIBentryALTinterwordstretchfactor\fontdimen3\font minus \fontdimen4\font\relax}
\providecommand{\BIBforeignlanguage}[2]{{%
\expandafter\ifx\csname l@#1\endcsname\relax
\typeout{** WARNING: IEEEtran.bst: No hyphenation pattern has been}%
\typeout{** loaded for the language `#1'. Using the pattern for}%
\typeout{** the default language instead.}%
\else
\language=\csname l@#1\endcsname
\fi
#2}}
\providecommand{\BIBdecl}{\relax}
\BIBdecl

\bibitem{MW1}
X.~Shao \emph{et~al.}, ``A tutorial on six-dimensional movable antenna for {6G} networks: Synergizing positionable and rotatable antennas,'' \emph{IEEE Commun. Surv. Tutorials}, vol.~28, pp. 3666--3709, Aug. 2025.

\bibitem{Ma11}
W.~Ma \emph{et~al.}, ``A survey on reconfigurable and movable antennas for wireless communications and sensing,'' \emph{IEEE Commun. Surv. Tutorials}, vol.~28, pp. 4842--4882, Feb. 2026.

\bibitem{Tan2022ArrayTopologies}
W.~Tan and S.~Ma, ``Antenna array topologies for mmwave massive {MIMO} systems: Spectral efficiency analysis,'' \emph{IEEE Trans. Veh. Technol.}, vol.~71, no.~12, pp. 12\,901--12\,915, Dec. 2022.

\bibitem{ZhouCong1}
C.~Zhou \emph{et~al.}, ``Sparse array enabled near-field communications: Beam pattern analysis and hybrid beamforming design,'' \emph{IEEE Trans. Wireless Commun.}, vol.~24, no.~12, pp. 10\,261--10\,277, Dec. 2025.

\bibitem{XIU4}
Y.~Xiu, Y.~Zhao, R.~Yang, D.~Niyato, J.~Jin, Q.~Wang, G.~Liu, and N.~Wei, ``Movable antenna-aided cooperative {ISAC} network with time synchronization error and imperfect {CSI},'' \emph{IEEE Trans. Commun.}, vol.~74, pp. 2968--2983, May 2025.

\bibitem{CLA7}
M.~Kurras \emph{et~al.}, ``On the application of cylindrical arrays for massive {MIMO} in cellular systems,'' in \emph{Proc. 22nd Int. ITG Workshop on Smart Antennas}, Mar. 2018, pp. 1--8.

\bibitem{Wu2023UCA}
Z.~Wu, M.~Cui, and L.~Dai, ``Enabling more users to benefit from near-field communications: From linear to circular array,'' \emph{IEEE Trans. Wireless Commun.}, vol.~23, no.~4, pp. 3735--3748, Apr. 2024.

\bibitem{Wu2024UCA}
Z.~Wu and L.~Dai, ``The manifestation of spatial wideband effect in circular array: From beam split to beam defocus,'' \emph{IEEE Trans. Commun.}, vol.~72, no.~5, pp. 3064--3078, May 2024.

\bibitem{ISAC_FCLA}
W.~Jiang, Z.~Wei, F.~Liu, Z.~Feng, and P.~Zhang, ``Collaborative precoding design for adjacent integrated sensing and communication base stations,'' \emph{IEEE Internet Things J.}, vol.~11, no.~9, pp. 15\,059--15\,074, May 2024.

\bibitem{CLA4}
D.~G. Riviello and F.~D. Stasio, ``{5G} beamforming implementation and trade-off investigation of cylindrical array arrangements,'' in \emph{Proc. Int. Symp. Wirel. Pers. Multimed. Commun. (WPMC)}, Nov. 2019, pp. 1--6.

\bibitem{CLA1}
E.~Yaacoub \emph{et~al.}, ``{3D} beamforming with massive cylindrical arrays for physical layer secure data transmission,'' \emph{IEEE Commun. Lett.}, vol.~23, no.~5, pp. 830--833, May 2019.

\bibitem{XIU5}
Y.~Xiu, Y.~Zhao, R.~Yang, H.~Tang, L.~Qu, M.~Khabbaz, C.~Assi, and N.~Wei, ``Latency minimization for movable relay-aided {D2D-MEC} communication systems,'' \emph{IEEE Trans. Mobile Comput.}, vol.~25, no.~1, pp. 533--549, Jan. 2026.

\bibitem{CLAS}
J.~Yang \emph{et~al.}, ``A cylindrical phased array radar system for {UAV} detection,'' in \emph{Proc. 6th Int. Conf. Intell. Comput. Signal Process. (ICSP)}, Apr. 2021, pp. 894--898.

\bibitem{MA1}
L.~Zhu, W.~Ma, and R.~Zhang, ``Modeling and performance analysis for movable antenna enabled wireless communications,'' \emph{IEEE Trans. Wireless Commun.}, vol.~23, no.~6, pp. 6234--6250, Jun. 2024.

\bibitem{zlp1}
------, ``Movable-antenna array enhanced beamforming: Achieving full array gain with null steering,'' \emph{IEEE Commun. Lett.}, vol.~27, no.~12, pp. 3340--3344, Dec. 2023.

\bibitem{zlp2}
L.~Zhu \emph{et~al.}, ``A tutorial on movable antennas for wireless networks,'' \emph{IEEE Commun. Surv. Tutorials}, vol.~28, pp. 3002--3054, Feb. 2026.

\bibitem{MEI2}
W.~Mei, X.~Wei, B.~Ning, Z.~Chen, and R.~Zhang, ``Movable-antenna position optimization: A graph-based approach,'' \emph{IEEE Wireless Commun. Lett.}, vol.~13, no.~7, pp. 1853--1857, Jul. 2024.

\bibitem{XIU1}
Y.~Xiu, W.~Lyu, Y.~Li, R.~Yang, P.~L. Yeoh, W.~Zhang, G.~Liu, and N.~Wei, ``Meta-reinforcement learning optimization for movable antenna-aided full-duplex {CF-DFRC} systems with carrier frequency offset,'' \emph{IEEE Trans. Commun.}, vol.~74, pp. 5803--5819, Mar. 2026.

\bibitem{XIU2}
Y.~Xiu, Y.~Zhao, K.~Wang, M.~Xu, D.~Niyato, G.~Liu, and N.~Wei, ``Delay minimization for movable antennas-enabled anti-jamming communications with mobile edge computing,'' \emph{IEEE Trans. Commun.}, vol.~74, pp. 6243--6258, Mar. 2026.

\bibitem{zlp4}
Y.~Ma \emph{et~al.}, ``Movable antenna-enhanced secure communication: Opportunities, challenges, and solutions,'' \emph{IEEE Wireless Commun.}, pp. 1--8, Early Access, 2025.

\bibitem{ZhouCong2}
C.~Zhou \emph{et~al.}, ``Frequency-switching array enhanced physical-layer security in terahertz bands: A movable antenna perspective,'' \emph{arXiv preprint arXiv:2507.01624}, Jul. 2025.

\bibitem{XieLei}
L.~{Xie}, P.~{Wang}, G.~{Shen}, G.~{Li}, W.~{Mei}, and L.~{Chen}, ``{Secure Communication in {MIMOME} Movable-Antenna Systems with Statistical Eavesdropper CSI},'' \emph{arXiv e-prints}, p. arXiv:2601.14755, Jan. 2026.

\bibitem{MEI3}
X.~Shen \emph{et~al.}, ``Movable-antenna-enhanced physical-layer service integration: Performance analysis and optimization,'' \emph{IEEE Wireless Commun. Lett.}, vol.~14, no.~9, pp. 2952--2956, Sep. 2025.

\bibitem{Ran1}
R.~{Yang} \emph{et~al.}, ``{Movable Antenna Empowered Covert Dual-Functional Radar-Communication},'' \emph{arXiv e-prints}, p. arXiv:2601.14868, Jan. 2026.

\bibitem{zlp3}
Y.~Ma, K.~Liu, Y.~Liu, L.~Zhu, and Z.~Xiao, ``Movable-antenna aided secure transmission for {RIS-ISAC} systems,'' \emph{IEEE Trans. Wireless Commun.}, vol.~24, no.~12, pp. 10\,019--10\,035, Dec. 2025.

\bibitem{XIU3}
Y.~Xiu, Y.~Zhao, R.~Yang, W.~Lyu, D.~Niyato, D.~In~Kim, G.~Liu, and N.~Wei, ``Robust optimization for movable antenna-aided cell-free {ISAC} with time synchronization errors,'' \emph{IEEE Trans. Wireless Commun.}, vol.~25, pp. 10\,082--10\,097, Jan. 2026.

\end{thebibliography}

\vspace{12pt}

\end{document}